\documentclass[preprint,12pt,english]{elsarticle}

\usepackage[T1]{fontenc}
\usepackage[latin9]{inputenc}
\usepackage{geometry}
\usepackage{booktabs}
\usepackage{url}
\usepackage{amsmath,amssymb,amsfonts}
\usepackage{graphicx}
\usepackage{color}
\usepackage[superscript,biblabel]{cite}
\usepackage{babel}
\usepackage{listings}
\usepackage{hyperref}
\usepackage{braket}

\newcommand\ee{\mathrm{e}}
\newcommand\ii{\mathrm{i}}
\newcommand\iv{\ii\nu}
\newcommand\ivp{\ii\nu^{\prime}}
\newcommand\iw{\ii\omega}
\newcommand\taup{\tau^{\prime}}
\newcommand\dd[1]{\mathrm{d}#1\:}

\newcommand\sigmap{\sigma^{\prime}}
\newcommand\sigmapp{\sigma^{\prime\prime}}
\newcommand\sigmappp{\sigma^{\prime\prime\prime}}
\newcommand\sigmapppp{\sigma^{\prime\prime\prime\prime}}
\newcommand\eye{\boldsymbol{1}}
\newcommand\tr{\operatorname{tr}}
\newcommand\sgn{\operatorname{sgn}}
\newcommand\rect{\operatorname{rect}}
\newcommand\cee{{\hat c}^{\protect\vphantom\dagger}}
\newcommand\cdag{{\hat c}^{\dagger}}
\newcommand\eff{{\hat f}^{\protect\vphantom\dagger}}
\newcommand\fdag{{\hat f}^{\dagger}}
\newcommand\up{\uparrow}
\newcommand\down{\downarrow}

\definecolor{verde}{rgb}{0.0,0.5,0.0}
\definecolor{mayablue}{rgb}{0.45, 0.76, 0.98}
\definecolor{purple}{rgb}{0.55,0.0,0.5}

\newcommand {\onlinecite}{\citenum}

\newcommand{\upfolded}[1]{\hat{\boldsymbol{#1}}}
\newcommand{\lcee}{{\hat{\boldsymbol c}^{\protect\vphantom\dagger}}}
\newcommand{\lcdag}{{\hat{\boldsymbol c}^\dagger}}

\newcommand\mn[1]{\text{#1}}
\newcommand\ek[1]{{\left[ #1 \right]}} 

\newcounter{bla}

\journal{Computer Physics Communications}

\begin{document}

\begin{frontmatter}

\title{w2dynamics: Local one- and two-particle quantities from dynamical mean field theory}

\author[a,b]{Markus Wallerberger\corref{corr}}
\ead{mwallerb@umich.edu}

\author[c]{Andreas Hausoel}
\author[a]{Patrik Gunacker}
\author[c]{Alexander Kowalski}
\author[c]{Nicolaus Parragh}
\author[c]{Florian Goth}
\author[a]{Karsten Held}

\author[c]{Giorgio Sangiovanni}
\ead{sangiovanni@physik.uni-wuerzburg.de}

\cortext[corr]{Corresponding author; user mailing list: \url{w2dynamics-users@list.tuwien.ac.at}}
\address[a]{Institute of Solid State Physics, TU Wien, 1040 Vienna, Austria}
\address[b]{Department of Physics, University of Michigan, Ann Arbor, MI 48109, USA}
\address[c]{Institut f\"ur Theoretische Physik und Astrophysik, Universit\"at W\"urzburg, D-97074 W\"urzburg, Germany}

\begin{abstract}
We describe the hybridization-expansion continuous-time quantum Monte Carlo 
code package ``w2dynamics'', developed in Wien and W\"urzburg. We discuss the main features of this  multi-orbital quantum impurity solver
for the Anderson impurity model, dynamical mean field theory as well as 
its coupling to density functional theory. The  w2dynamics package allows for  calculating one- and two-particle quantities; it includes worm and further novel sampling schemes. Details about its download, installation, functioning and the relevant parameters are provided.

\end{abstract}


\end{frontmatter}


{\bf PROGRAM SUMMARY}

\begin{small}
\noindent
{\em Program Title:}  w2dynamics                                       \\
{\em Licensing provisions:} GNU General Public License (GPLv3)         \\
{\em Programming language:} Python, Fortran 90, and C++11              \\
{\em Required dependencies:} cmake ($\ge 2.8.5$), MPI, LAPACK, FFTW3, Python ($\ge 2.4$) \\
{\em Optional dependencies:}
    NFFT, pip, numpy ($\ge 1.4$), scipy ($\ge 0.10$), h5py, mpi4py, configobj

{\em Nature of problem:}\\
   Numerically unbiased solutions of one- and two-particle propagators
   for quantum impurity models at finite temperature.  Approximate solutions
   for general lattice models with strong electronic correlation.

{\em Solution method:}\\
   Continuous-time quantum Monte Carlo in the hybridization expansion,
   including worm sampling, for the impurity problem.  Dynamical mean field
   theory solver for the lattice problem.

\end{small}

\section{Introduction}

Strongly correlated electron systems exhibit a range of fascinating
phenomena such as the Mott-Hubbard metal-to-insulator transition,
spin- and charge-density wave states, giant magnetoresistance, and
heavy fermion behavior. However even a numerical solution of correlated lattice models, not to speak of realistic materials calculations, is impossible in practice, except for very few lattice sites,  because the Fock space scales exponentially with the number of lattice sites and orbitals. The single impurity Anderson model (SIAM) \cite{Hewson1993} is the rare exception of a correlated electron model that is non-trivial while a numerical solution is still feasible.
For a single (or a few) orbital(s) this can be done by the the numerical renormalization group (NRG) \cite{Bulla2008}, matrix product states (MPS) \cite{Schollwock2005,Ganahl2014,Wolf2015}, or, if the conduction electron bath is discretized by a few sites, by the exact diagonalization.
When more orbitals are taken into account or  for calculating general two-particle Green's functions with three independent frequencies, continuous-time quantum Monte Carlo  (CT-QMC) approaches \cite{Rubtsov2005,gull-rmp-2011} are the method of choice.

The core of the w2dynamics package is a CT-QMC 
calculation of  $n$-particle Green's functions for a  generalized SIAM
using the  hybridization expansion  (CT-HYB) \cite{Werner2006}; closely related codes are  ALPS \cite{ALPS2,ALPSCORE}, TRIQS \cite{TRIQS}, and   EDMFTF~\cite{Haule2007}.
The CT-HYB approach
 expands the partition function in terms of the bath hybridization and stochastically samples the resulting
determinants of the diagrammatic series.
We  employ the matrix-vector technique for the local time evolution \cite{gull-rmp-2011}. Furthermore  worm sampling is used to compute  most general 
 two-particle Green's functions \cite{gunacker-prb-2015} and  improved estimators are used 
for the self-energy \cite{gunacker-prb-2016} which help reduce the CT-HYB's signal-to-noise ratio which is  particularly problematic at  high frequencies.\cite{gunacker-prb-2016}

The CT-HYB algorithm of w2dynamics might serve directly to study quantum dots connected to non-interacting leads or magnetic impurities in weakly interacting solids.
Dynamical mean field theory (DMFT)  \cite{georges-rmp-1996}, on the other hand, maps  Hubbard-like lattice Hamiltonians onto the self-consistent solution of a SIAM. Indeed a major application of  w2dynamics is the  solution of lattice Hamiltonians within DMFT. These might be model systems
or, for materials calculations,  Hamiltonians derived from {\em ab initio}
calculations such as  density functional theory (DFT) \cite{Martin04,Gunnarsson2003}  or the  $GW$ approximation\cite{Aryasetiawan1998}.

The most widely  approach of the latter kind is DFT+DMFT\cite{anisimov-jpcm-1997,Lichtenstein1998,kotliar-rmp-2006,Held2007}, others approaches include 
  $GW$+DMFT \cite{Sun02,Biermann2003,janQSGWDMFT} and more general
embedding schemes \cite{knizia-prl-2012,lan-jcp-2015}.
All these  methods first attempt
to identify the correlated subspace, map that space onto an effective
model Hamiltonian by estimating the model parameters, 
correct for double counted contributions  already included in the \emph{ab-initio } calculation, and finally
find an (approximate) solution for this {\em ab initio}-derived model Hamiltonian.
The last step is provided by w2dynamics which includes  an interface to 
wannier90 \cite{Mostofi2008} generated Hamiltonians, obtained e.g.\
from Wien2k \cite{Schwarz2002} through the wien2wannier interface \cite{Kuneifmmodecheckselsevsfi2010a}.
Depending on the method employed, the w2dynamics results  of the subspace might need to be embedded again into the 
full  \emph{ab-initio} space, and one might 
(optionally) iterate the whole procedure until self-consistency. \cite{Savrasov2004,Aichhorn2011b,Bhandary2016}

While DMFT is restricted to site-local correlations, post-DMFT methods aim at including non-local correlations on top of DMFT. One route to this end are cluster extensions \cite{maier-rmp-2005} of DMFT which consider a cluster of sites within a DMFT-like bath. In principle, w2dynamics allows for such calculations by considering the whole cluster as the impurity, but with CT-HYB one is faced with a severe sign problem already at moderate cluster sizes.  Other CT-QMC methods fare better in this case, but are restricted to single orbitals and/or density-density-like interactions.\cite{gull-rmp-2011}

Another route are diagrammatic extensions of DMFT. \cite{toschi-prb-2007,rubtsov-prb-2008,georgPRB2012,taranto-prl-2014,rohringer-prb-2013,valliPRB2015,Li2015,ayral-prb-2016,ayral-prb-2016b,RMPVertex}
These require the calculation of the local two-particle vertex which depends on three frequencies and can be
calculated by w2dynamics; simplified variants \cite{ayral-prb-2016} are based on three-leg (two-frequency) vertices while at an higher order in the expansion
around DMFT (or for an error estimate) also three-particle vertices are required.  \cite{Ribic2017b}
 One of the strong points of w2dynamics is the
 calculation of general multi-orbital two-particle  Green's functions, which are needed 
to obtain the  local vertices of post-DMFT approaches and for calculating DMFT susceptibilities. This is based on an intensive code and algorithmic development, including worm sampling, \cite{gunacker-prb-2015} improved estimators, \cite{gunacker-prb-2016} and vertex asymptotics \cite{Kaufmann2017};
w2dynamics can also be employed to calculate selected three-particle Green's functions.  \cite{Ribic2017b}
From the  local vertex calculated by w2dynamics, the diagrammatic extensions of DMFT determine in turn non-local vertices and self-energies, using ladder or parquet diagrams. \cite{RMPVertex}

The outline of the paper is as follows:
A description of the code is provided in Sections~\ref{sec:dmft} and \ref{sec:cthyb}. Here, Section~\ref{sec:dmft} provides an overview of the Python scripts for doing self-consistent DMFT calculations of various types, which wrap around the CT-HYB code.  Section~\ref{sec:cthyb} in turn outlines the
CT-HYB code for solving the SIAM. This includes the Monte Carlo steps of
e.g.\ inserting and removing hybridization lines in  Section \ref{Sec:Step},
routines for estimating physical observables in Section \ref{Sec:Estimator},
and the calculation of the local and bath weight in Section  \ref{Sec:LocalTrace} and  \ref{Sec:BathTrace}, respectively.
Section \ref{Sec:Tutorial} provides a brief description on how to
install and run the code; and Section \ref{sec:io}  describes which parameters the user can (has to) set and the format of the hdf5 data output. 
These latter two, user-oriented  Sections only provide a very first introduction, for further details we refer the reader to the online tutorial and wiki
(cf.~Section \ref{Sec:Where}). Finally, Section \ref{Sec:conclusion} provides a conclusion.

\section{Dynamical mean field theory (DMFT) solver\label{sec:dmft}}

\begin{figure}
\includegraphics[width=1\textwidth]{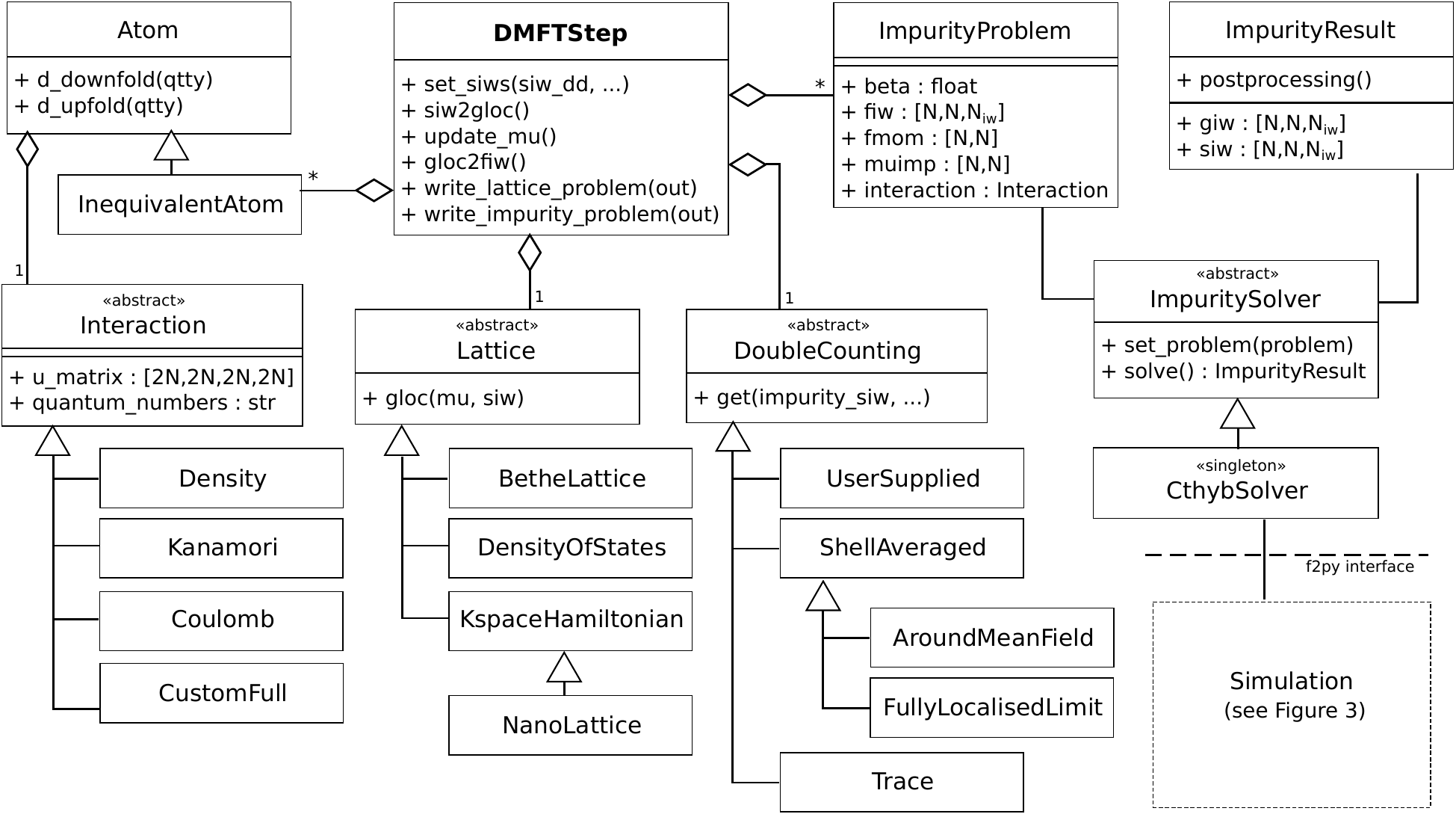}
\caption{Simplified  unified modeling language (UML) class diagram of the Python structures relevant to
the DMFT self-consistency loop. The triangle denotes a subclass relation; the diamond an aggregation, i.e. that one object ``has an'' other object. All base classes reside in a Python
module with the same name in the \textsf{dmft} directory (except for classes beginning with \textsf{Impurity}, which all reside in \textsf{impurity.py}).}
\label{fig:uml-python}
\end{figure}

Before turning to the w2dynamics solution of a SIAM, let us discuss the Python
scripts that allow for an outer DMFT loop.
This DMFT treatment allows us to  deal with  Hubbard-like lattice model,  which in full generality can be written as:
\begin{equation}\begin{split}
  \upfolded{H}_\mathrm{int} &=
    \sum_{R,R^\prime} \sum_{I,J} \sum_{\mu\nu} \upfolded{H}^{(IJ)}_{\mu,\nu}(R,R^\prime)
   \; \lcdag{}^{(I)}_\mu(R) \; \lcee{}^{(J)}_\nu(R^\prime)\\ 
  &+ \frac12 \sum_R \sum_I \sum_{\kappa\lambda\mu\nu} U^{(I)}_{\kappa\lambda\mu\nu}
   \; \lcdag{}^{(I)}_\kappa(R) \; \lcdag{}^{(I)}_\lambda(R) \; \lcee{}^{(I)}_\nu(R) \; \lcee{}^{(I)}_\mu(R)\; . 
\end{split}
\label{hlatt}
\end{equation}
Here,  $\lcee{}^{(I)}_\mu(R)$ annihilates a fermion, $R$ denotes the coordinates of
a unit cell or supercell, $I$ denotes the atom index within a supercell, and $\mu$
runs over spin-orbitals.  $\upfolded{H}$ parameterizes the quadratic or non-interacting
term, while $U$ parameterizes the interaction, which is assumed to be local for each
atom.  With bold typeface we indicate full matrices in the basis of all ``correlated'' and ``ligand'' spin-orbitals
(included in the low-energy subspace) of all atoms of the unit cell;
while non-bold-typeface tensors are restricted to the  ``correlated''
space only as e.g.\ the Coulomb interaction.
The small hat-symbol indicates that the quantity is a vector or a matrix, even if restricted to one of the atoms of the unit cell. The corresponding basis is defined by all spin-orbitals of the low-energy subspace associated to that atom.

The DMFT code is implemented in Python, since the challenges there
are mostly conceptual, and Python allows for a more compact and
clearer implementation. Most of the numerical problems are standard
matrix operations as well as fast Fourier transforms, which can be
deferred to the numpy package at essentially no loss of computing
speed. For the CT-QMC solver (see Section~\ref{sec:cthyb}), this
is not a viable strategy. Instead, it has been implemented in Fortran
90, and the interface is mediated by the f2py interface generator
included with numpy.

The different parts of the DMFT loop were carefully modularized
in order to allow for their effective unit testing and re-use in different
codes. A high-level UML class diagram is shown in Figure~\ref{fig:uml-python},
which is centered around the ``driver class'' \textsf{DMFTStep}.
In the following, we will explain the working of the code by going
through the different modules in the code.

\paragraph{Lattice}

This module defines an abstract class  \textsf{Lattice}  for  
the computation of the local Green's function $\upfolded{G}_\text{loc}(\iv)$.

With bold typeface we indicate again all quantities within the full basis of all ``correlated'' and ``ligand'' spin-orbitals
(included in the low-energy subspace) of all atoms of the unit cell.
The  local Green's function is calculated from the ${k}$-integrated Dyson equation for the lattice model
\begin{align} \label{Gloc}
\upfolded{G}_{\mathrm{loc}}(\iv) =
\frac{1}{N_k} \sum_{k}
\left[ (\iv + \mu) \upfolded{\eye} - \upfolded{H}(k) - \upfolded{\Sigma}(\iv) -\upfolded{\Sigma}_{\mathrm{DC}} \right]^{-1}. 
\end{align}
This is the definition used 
by setting \textsf{DOS=ReadIn},
        the definition Eq.(\ref{Gloc}) is directly realized using 
$\upfolded{H}(k)$   (read in from \textsf{HkFile}) on the given grid of $k$-points.
With \textsf{DOS=Bethe} the code replaces, for the special case of a Bethe lattice, the $k$-summation by an integral over a semicircular density of states.
\textsf{DOS=nano} gives the possibility of adding frequency-dependent functions describing the (physical) hybridization to non-interacting external leads, see  Refs.~\onlinecite{valli-prl-2010,Schueler2017}. 

The code supports calculations at a fixed total filling  $N_{\mathrm{target}}$
by adjusting
the chemical potential to $N(\mu)=N_{\mathrm{target}}$. For a
$k$-discretized Hamiltonian this root finding  for $\mu$  involves a series of local Green's
function computation at different $\mu$. This may easily become the bottleneck
of the code for large systems since it scales as $O(N_{\mathrm{fl}}^{3}N_{k}N_{\omega}N_{\mathrm{it}})$,
where $N_{\mathrm{fl}}$ is the number of spin-orbitals, $N_{k}$
is the number of $k$-points, $N_{\omega}$ is the number of frequencies,
and $N_{\mathrm{it}}$ is the number of iterations in the root finder.
One can however pre-compute the eigenvalues of $(\upfolded{H}-\upfolded{\Sigma})$, store
them as $\eta_{i}(k,\iv)$, and rewrite the density as:
\begin{equation}
N(\mu)=\frac{1}{\beta}\sum_{\nu}\ee^{-\iv0^{-}}\tr \upfolded{G}_{\mathrm{loc}}(\iv)
=\frac{1}{\beta N_k}\sum_{k,\nu}\sum_{i=1}^{N_{\mathrm{fl}}}\frac{\ee^{-\iv0^{-}}}{\iv+\mu-\eta_{i}(k,\iv)}.
\end{equation}
This is a CPU\textendash memory tradeoff as it reduces the scaling to $O(N_{\mathrm{fl}}^{3}N_{k}N_{\omega})+O(N_{\mathrm{it}}N_{\mathrm{fl}}N_{k}N_{\omega})$
while requiring $O(N_{\mathrm{fl}}N_{k}N_{\omega})$ of memory. If
not enough memory is available, the code falls back to the direct strategy.

The lattice code itself is (trivially) parallelized over frequency
since instead of $N(\mu)$ the trace over $\upfolded{G}_{\mathrm{loc}}(\iv)$
and the corresponding model is computed, which can then be summed
at a higher level.

\begin{figure}
\begin{centering}
\includegraphics{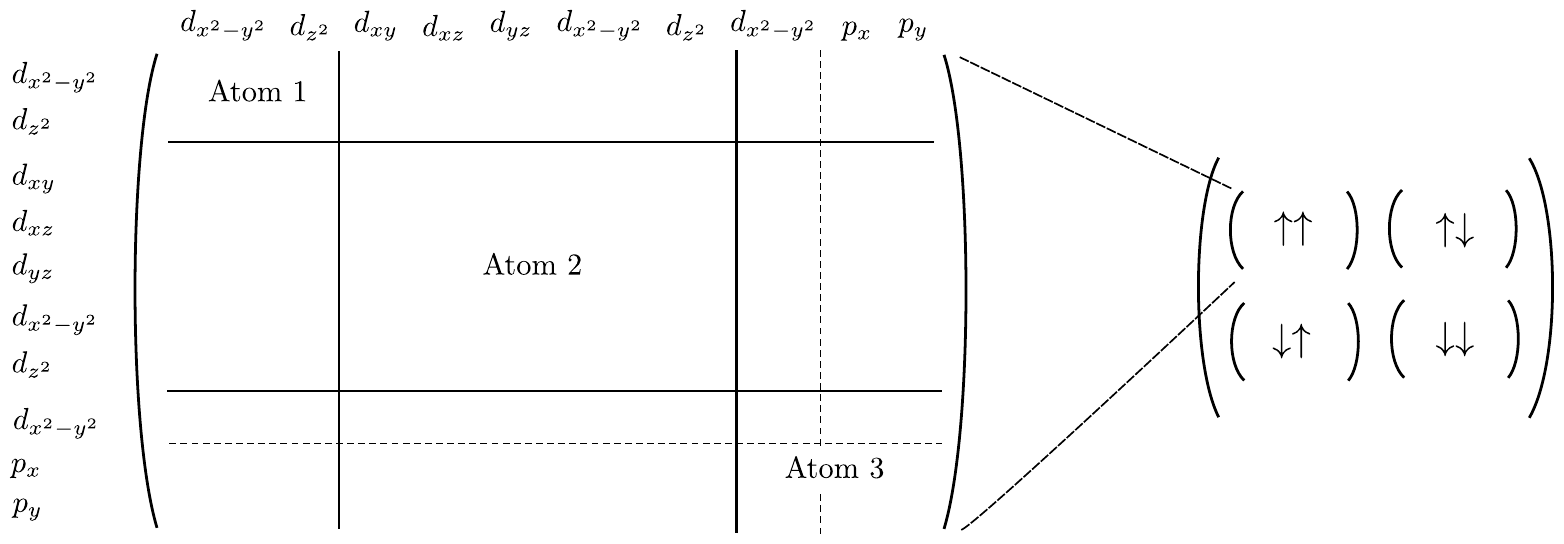}
\par\end{centering}
\caption{Example of the atom, spin and orbital structure of the single-particle Hamiltonian $\upfolded{H}(k)$ (at a fixed $k$-point), input for \textsf{DOS=ReadIn}. The unit cell of this example contains three atoms, the first one with $e_g$-orbitals only, the second with the full $d$-shell and the third with a $x^2-y^2$-orbital and two ligands (called here ``$p$''). Each element of this Hamiltonian matrix has to have a real and an imaginary part. The only syntax requirement for the DMFT loop to work correctly is that for each atom, the ``correlated'' orbitals (in this case the $d$ ones) have to come before the ``uncorrelated'' ones.}
\label{fig:atom-structure}
\end{figure}

\paragraph{Atom}

There are several reasons why realistic calculations of materials often require several ``correlated'' atoms, $I=1,\ldots,N_\mathrm{at}$, in the unit cell:  among others magnetic supercells, heterostructures and surfaces. 
To some extent, the choice of which orbitals belong to the same atom is
arbitrary and it lies in the  responsibility of the user that the
grouping into ``atoms'' is consistent with the (local) interaction parameters
used between the orbitals.

In  single-site DMFT,  there are only intra-atomic self-energies and no self-energies between different atoms \cite{pothoff-prb-1999,florens-prl-2007,snoek-njphys-2008,valli-prl-2010}. 
$\upfolded{\Sigma}(\iv)$ is therefore an atom-block-diagonal matrix in the basis sketched in Fig.~\ref{fig:atom-structure}, whereas $\upfolded{H}(k)$ contains instead  hoppings between atoms as well as intra-atomic local and non-local elements. 
One can go from the full basis of all $N_\mathrm{fl}$   ``flavors'' within the supercell (i.e., atom, spin and orbital flavor; full matrix as sketched in Fig. \ref{fig:atom-structure}) to the block of a given atom $I$ with  $N_\mathrm{fl}^I$ flavors (spin and orbital; block Atom $I$ in  Fig. \ref{fig:atom-structure}):
\begin{subequations}
\begin{align}
\hat{G}^{(I)}(\iv) & = D_{I}\upfolded{G}_{\mathrm{loc}}(\iv)D_{I}^{\rm tr}\label{eq:downfold-1}\\
\upfolded{\Sigma}(\iv) & ={\textstyle \sum}_{I=1}^{N_{\mathrm{at}}}D^{\rm tr}_{I}\hat{\Sigma}^{(I)}(\iv)D_{I}\label{eq:upfold-1},
\end{align}
\label{eqs:updownfold}\end{subequations}
where $D_{I}$ is the ${N_\mathrm{fl}}^{(I)}\times N_\mathrm{fl}$ matrix
that projects (``downfolds'') to the impurity problem of the $I$-th atom with ${N_\mathrm{fl}}^{(I)}$
local degrees of freedom; 
$\hat{G}^{(I)}$ and $\hat{\Sigma}^{(I)}$ are the corresponding local Green's function and self-energy matrices, respectively.
In our case, $D_{I}$ just projects out some rows and columns of the lattice
problem (Fig.~\ref{fig:atom-structure}). 
The transposed $N_\mathrm{fl}\times{N_\mathrm{fl}}^{(I)}$  matrix $D_{I}^{tr}$
allows for an extension (``upfolding'') of atom $I$-based quantities to the full supercell of all atoms through Eq.~(\ref{eq:upfold-1}), and satisfies
$D_{I}D_{J}^{tr}=\delta_{IJ}\eye^{(I)}$. The up- and down-folding procedure
is handled by the \textsf{Atom} class.

What we call here ``$p$''- or ligand-orbitals are degrees of freedom that enter the low-energy problem and can hence exchange electrons with the ``correlated'' subspace of the ``$d$''-orbitals. These  ``$p$''-orbitals are treated as non-interacting, have a vanishing self-energy and do not need to be included in the CT-QMC for the SIAM.  The code can however also take care of an explicit Hartree contribution by setting $U_{pp}$ and $U_{pd}$ kind of interactions between the  ``$p$''-orbitals and between  ``$p$''- and ``$d$''-orbitals respectively. \cite{hansmann-prl-2010,han-prl-2011,hansmann-njp-2014}. Please note that also the 
double counting becomes important if  both, ``$p$''- and ``$d$''-orbitals are included; it may also include 
 a contribution from an Hartree-like frequency-independent self-energy associated to these ``spectator'' degrees of freedom. 

The code can take local equivalence of atoms into account to minimize the
number of impurity problems that have to be solved (class \textsf{InequivalentAtom}).
If the parameter \textsf{EPSEQ} (see Sec.~\ref{sec:io}) is set, the code will
attempt to detect whether two atoms are equivalent by comparing their $G_{0}^{-1}$
matrices and the interaction $U$. The user can either tune \textsf{EPSEQ} or set its own equivalence pattern.
Magnetic biases given in the initial configuration (\textsf{se-shift}) as well as different interaction parameters also influence 
the equivalence pattern.

Both considering  the  ``$p$''-orbitals as non-interacting as well as ignoring correlation between atoms inside a super-cell allows one to 
work around the exponential scaling of the impurity solver with the number of impurity orbitals. Let us note here, that as a matter of course 
the user can also include  all atoms of the supercell into one 
w2dynamics ``atom''. In this case also correlations and self-energy elements between the physical atoms are included, at the expense of the increasing computational burden.

\begin{table}
\begin{centering}
\sf\begin{tabular}{llll}
\toprule
Hamiltonian= & Parameters & $H_\mathrm{int}=$ & conserves \\
\midrule
Density
& Udd, Jdd, Vdd
& $ H_\mathrm{dd}= \sum_i U n_{i\up} n_{j\down} +
\sum_{i<j}^{\sigma,\sigmap} (V - J\delta_{\sigma\sigmap}) n_{i\sigma} n_{j\sigmap}
$
& $N$, $S_z$, $n_{i\sigma}$
\\ \addlinespace[0.4em]
Kanamori
& Udd, Jdd, Vdd
& $ H_\mathrm{dd} - J\sum_{i<j} (\cdag_{i\up} \cdag_{i\down} \cee_{j\up} \cee_{j\down}
  + \cdag_{i\up} \cdag_{j\down} \cee_{j\up} \cee_{i\down})
$
& $N$, $S_z$, $\mathrm{PS}$
\\ \addlinespace[0.4em]
Coulomb
& F0, F2, F4
& $\sum_n F_n \ldots$
& $N$, $S_z$, $L_z$
\\ \addlinespace[0.4em]
ReadUmatrix
& Umatrix
& $
   \frac12 \sum_{ijkl} U_{ijkl} \sum_{\sigma\sigmap}
        \cdag_{i\sigma} \cdag_{j\sigmap} \cee_{l\sigmap} \cee_{k\sigma}
$
& $N$, $S_z$
\\ \addlinespace[0.4em]
\bottomrule
\end{tabular}
\par\end{centering}
\caption{Available parametrizations of the local interaction. $\cee_{i\sigma}$
   annihilates a fermion on the corresponding impurity, $n_{i\sigma}=\cdag_{i\sigma}\cee_{i\sigma}$
    is the corresponding density operator, $i,j,k,l$ are impurity
   orbitals and $\sigma,\sigma^\prime$ denote spins. $N=\sum_i(n_{i\uparrow}+n_{i\downarrow})$
   is the total occupation, $S_z=\sum_i(n_{i\uparrow}-n_{i\downarrow})$ the total spin in $z$ direction and $\mathrm{PS}$
   denotes the pattern of singly-occupied orbitals (see Ref. \onlinecite{parragh-prb-2012} for a definition).  Note that the quantities conserved by the interaction given here may still be broken by the one-particle (non-interacting) part of the local Hamiltonian".
}
\label{tab:int}
\end{table}

\paragraph{Local interaction}

The code supports different parameterization of the local interaction
Hamiltonian, summarized in Table~\ref{tab:int}.

These can be selected by choosing the parameter with the name \textsf{Hamiltonian} (capable of being misunderstood),
in the \textsf{[Atoms]} section.
The simplest one is the ``density-density'' scheme (\textsf{Density}), where only two-body interactions that can be written in terms of orbital occupations appear. The couplings of this kind of interaction are set via the parameters \textsf{Udd}, representing the local Hubbard repulsion, the Hund coupling \textsf{Jdd} and \textsf{Vdd}, which need not fulfill specific relation with the other two parameters. Note that the inclusion of \textsf{Jdd} in the density-density scheme breaks $\mathrm{SU(2)}$ symmetry, since spin flip and pair hopping terms cannot be cast in this form. 
These parameters are used uniformly for all orbitals of the given atom.  An orbital-dependent density-density type interaction can be set by using a special the \textsf{ReadNormalUmatrix} scheme.

\textsf{Hamiltonian=Kanamori} adds the pair-hopping and spinflip terms to the density-density interaction, thus restoring the $\mathrm{SU(2)}$ symmetry.\cite{georgesRevHund}.
This interaction scheme conserves the so-called ``PS''-quantum number,\cite{PS-number} which allows for a tighter block structure of the local Hamiltonian and thus effecient sampling.

With \textsf{Hamiltonian=Coulomb} the interaction matrix for a complete $d$-shell is generated from the Slater parameters $F_0$, $F_2$ and $F_4$ in spherical harmonics, and then transformed to the crystal field basis.

The other possibility is to use \textsf{ReadNormalUmatrix} in which a tensor of four orbital indices is expected as input  (\textsf{umatrix="filename"}). 
This tensor represents a $U_{ijkl}$ full Coulomb tensor, e.g. coming either from spherical Coulomb tensors written in the basis of the cubic crystal field eigenfunctions or directly from constraint random phase approximation (cRPA). 
\textsf{ReadUmatrix} reads a tensor with four orbital and four spin indices. 
Reading in the $U_{ijkl}$ tensor, the user has still the possibility of neglecting elements with a specific structure, for instance removing all entries that are not of the Kanamori type. This may be useful, for instance, in Kanamori calculations of the full $d$-shell, for which one however wants to keep the difference between intra-$t_{2g}$ and intra-$e_g$ Hund's couplings.
Let us note that if this is used in combination with ``$d$-$p$'' calculations, the user is advised that the double-counting values have to be set by hand rather than relying on the ``automatic'' calculation by w2dynamics.

For general interactions, the code automatically finds the structure of the local part of the impurity Hamiltonian with the minimal block size by adding \textsf{All} to the \textsf{QuantumNumber} list (cf. Sec.~\ref{sec:qn}).\cite{Parragh2013}

In the case of density-density type of local interactions, the code will support the addition of a retarded interaction term
$W^{(I)}$ by setting the \textsf{Screening} parameter.  This modifies
the effective impurity action as follows:\cite{DanielAndiUw}
\begin{equation}
  S^{(I)}[c,c^*] \to S^{(I)}[c,c^*]
    + \frac12 \sum_{\mu\nu} \int \dd{^2\tau} c^*_\mu(\tau) c_\mu(\tau) W^{(I)}_{\mu\nu}(\tau - \taup) c^*_\nu(\taup) c_\nu(\taup).
\end{equation}

\paragraph{ImpurityProblem}

At each step of the DMFT loop, $N_\mathrm{at}$ impurity problems are generated from the local Green's function of Eq.~(\ref{Gloc}).
The code calculates, for each atom $I$, the following Weiss field matrix (cf. Refs. \onlinecite{Georges1996,Kotliar2006,Held2007}):
\begin{align} \label{Weiss}
\left( \hat{\mathcal{G}}^{(I)}(\iv) \right)^{-1} = \left( D_I \upfolded{G}(\iv) D^{-1}_I \right)^{-1} + \hat{\Sigma}^{(I)}(\iv),
\end{align}
where $\upfolded{G}$ again denotes the full local Green's function of dimension $N_\mathrm{fl} \times N_\mathrm{fl}$, while $\hat\Sigma^{(I)}$ and ${\hat{\mathcal G}}^{(I)}$ denote
the self-energy and Weiss field, respectively, of the $I$'th atom and are of dimension $N_\mathrm{fl}^{(I)} \times N_\mathrm{fl}^{(I)}$. 
These Weiss fields, or the hybridization functions ${\Delta}$ related through Eq.~(\ref{G0hyb}),
 define  $N_\mathrm{at}$ auxiliary single-site quantum many-body impurity problems. Each inequivalent impurity problem  is then solved by \textsf{ImpuritySolver.solve()},  yielding the two-point Green's function of the $I$-th atom $\hat{G}^{(I)}_{\text{imp}}(\iv)$.
The corresponding self-energy is obtained via the Dyson equation
\begin{align}
\label{sigma}
\hat{\Sigma}^{(I)}(\iv) = \left( \hat{\mathcal{G}}^{(I)}(\iv) \right)^{-1}  - \left( \hat{G}_{\text{imp}}^{(I)}(\iv) \right)^{-1}
\end{align}
Once the self-energy matrices for all atoms are calculated via Eq.~(\ref{sigma}), the full matrix $\upfolded{\Sigma}(\iv)$ is constructed block-wise and inserted back into Eq.~(\ref{Gloc}). This way \textsf{ImpurityResult} generates a new local Green's function for the whole system. The self-consistency loop then goes on by calculating Eq.~(\ref{Weiss}) for each site again etc. The user is responsible to decide when convergence is reached.
To help the stabilization of the DMFT self-consistency loop we mix the previous and the current self-energies, according to the parameter \textsf{mixing} (a value of 0 means that the solution has no contribution from the previous iteration, whereas 1 means that the self-energy is entirely given by that at the previous iteration).  

\section{CT-HYB impurity solver}
\label{sec:cthyb}
The hybridization expansion continuous-time quantum Monte Carlo (CT-HYB) solver
provides an unbiased solution to  Anderson impurity
models arising e.g.\ from the self-consistent solution of the DMFT equations. 
The impurity Hamiltonian, in its full generality, can
be written as:
\begin{equation}\begin{split}
  H & = H_\mathrm{loc}[\cee, \cdag] + H_\mathrm{hyb}[\cee, \cdag, \eff, \fdag] + H_\mathrm{bath}[\eff, \fdag] \\
   & =
   \sum_{\kappa\lambda} E_{\kappa\lambda}\cdag_\kappa \cee_\lambda
   + \frac12 \sum_{\kappa\lambda\mu\nu} U_{\kappa\lambda\mu\nu} \cdag_\kappa \cdag_\lambda \cee_\nu \cee_\mu
   + \sum_{\kappa p} (V_{p\kappa} \fdag_p \cee_\kappa + \mathrm{h.c.}) + \sum_p \tilde E_p \fdag_p \eff_p,
\end{split}
\label{Eq:action}
\end{equation}
where $\cee_\lambda$ annihilates a fermion of spin-orbital $\lambda$ on the
impurity and $\eff_p$ annihilates a fermion on the bath, where the quantum number
$p$ can be continuous.  For the $I$-th impurity problem, the interaction tensor is $U = U^{(I)}$, and the quadratic part is given by
\begin{equation}
  (\hat{\mathcal{G}}^{(I)})^{-1}_{\lambda\mu}(\iv) = (\iv + \mu)\delta_{\lambda\mu} - E_{\lambda\mu} - \hat{\Delta}^{(I)}_{\lambda\mu}(\iv)
  \label{G0hyb}
\end{equation}
with the hybridisation function
\begin{equation}
  \hat{\Delta}^{(I)}_{\lambda\mu}(\iv) = \sum_p \frac {V^*_{\lambda p} V_{p\mu}}{\iv - \tilde E_p}.
  \label{hyb}
\end{equation}

\begin{figure}
\includegraphics[width=1\textwidth]{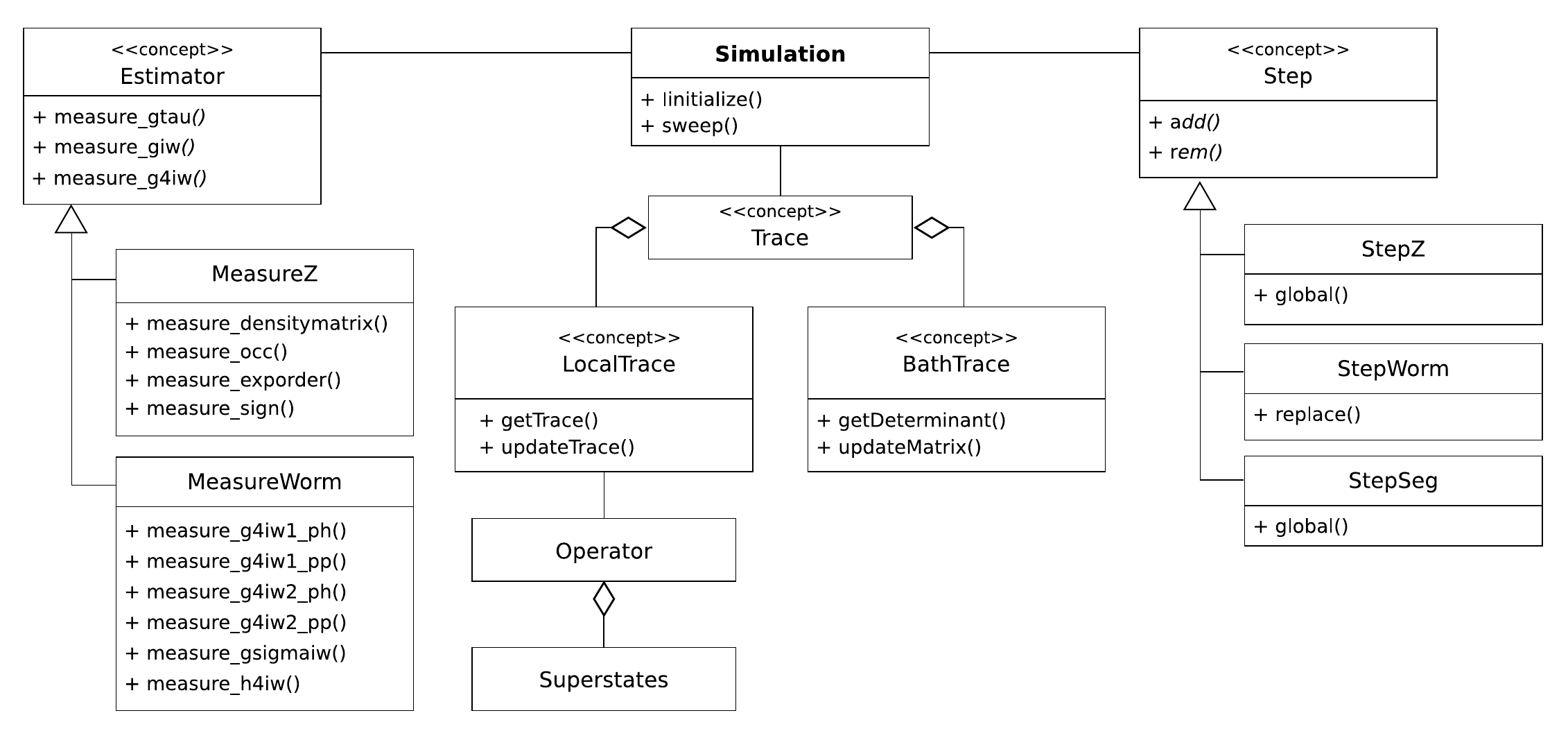}

\caption{Simplified UML diagram of the Fortran structures relevant to the CT-HYB
solver. <\textcompwordmark{}<concept>\textcompwordmark{}> marks logical
units that are technically not represented by Fortran structures.}

\label{fig:uml-qmc}
\end{figure}

CT-HYB expands the partition function $Z$ of that SIAM, defined by $H$ and the
inverse temperature $\beta$, in terms of the hybridisation $H_\mathrm{hyb}$
(for a derivation, see Ref.~\onlinecite{gull-rmp-2011}): 
\begin{equation}\begin{split}
   Z&= \sum_{k=0}^\infty  \int_0^\beta \prod_{i=1}^k \dd{\tau_i} \dd{\taup_i} \sum_{\lambda_i \lambda_i'}
       \tr \left[ T_\tau \mn{e}^{-\beta H_{\mn{loc}}} \prod_{i=1}^k  \cdag_{\lambda_i}(\tau_i) \cee_{\lambda_i'}(\tau_i') \right]
       \det \left[ \Delta_{\lambda_i \lambda_j} (\tau_i - \taup_j) \right]_{ij} \;.
\end{split}
       \label{eq_hybridisation_expansion}
\end{equation}
Here $T_\tau$ is the time-ordering operator, $H_{\mathrm{loc}}$ the local Hamiltonian of the impurity, cf. Eq.~(\ref{Eq:action})
and $\Delta$ is the hybridisation function, cf. Eq.~(\ref{hyb}). CT-HYB thus corresponds to the sampling over all diagrams with
vertices at ${\lambda_i,\lambda^\prime_i, \tau_i, \tau^\prime_i}$, and the weight of each diagram is given by the product of a
local weight, $w_\mathrm{loc}$, and a bath weight, $w_\mathrm{bath}$. Here, the local weight or local trace is given by
\begin{equation}
   w_{\mn{loc}} =  \mn{Tr} \ek{T_\tau \mn{e}^{-\beta H_{\mn{loc}}} \prod_{i=1}^k  d_{\lambda_i}^\dagger(\tau_i) d_{\lambda_i'}(\tau_i') },
   \label{g_loc_trace}
\end{equation}
which is represented by \textsf{LocalTrace} in the UML Figure \ref{fig:uml-qmc} and will be discussed in detail in Section \ref{Sec:LocalTrace}. The bath part of the weight ${w_{\mathrm{bath}}= \det\Delta}$  (\textsf{BathTrace} in Figure \ref{fig:uml-qmc}) is the determinant of noninteracting hybridization functions, whose calculation will be detailed in  Section  \ref{Sec:BathTrace}.
The Monte Carlo importance sampling of different flavors $\lambda_i$, imaginary time positions $\tau_i$ and expansion orders $k$ in
the sum of Eq.~(\ref{eq_hybridisation_expansion}) is realized through \textsf{Steps}
as outlined in  Section \ref{Sec:Step}. The Monte Carlo measurement of observables, on the other hand, is  performed by \textsf{Estimator} as discussed in Section \ref{Sec:Estimator}. 

The CT-QMC solver is written in Fortran 90 in order to account for
numerical efficiency when sampling the Monte Carlo configuration space.
Since Monte Carlo algorithms are  trivially parallelizable, we offset
the communication and averaging work to the Python infrastructure.
Consequently, the Fortran modules are entirely unaware of any parallelization
schemes.

The impurity solver is structured into several layers arranged in
order of increasing complexity. On the lowest level this includes
matrix operations and the random number generator. The intermediate
level includes routines for defining many-body operators and states.
These are used in the definition of the local problem and the bath
problem. On the highest level, the simulation block governs the Metropolis
Hastings Monte Carlo procedure. Monte Carlo steps and estimators are
defined for the partition function space and the worm space.

\subsection{Steps}
\label{Sec:Step}

\begin{table}[t]
\begin{centering}
\includegraphics{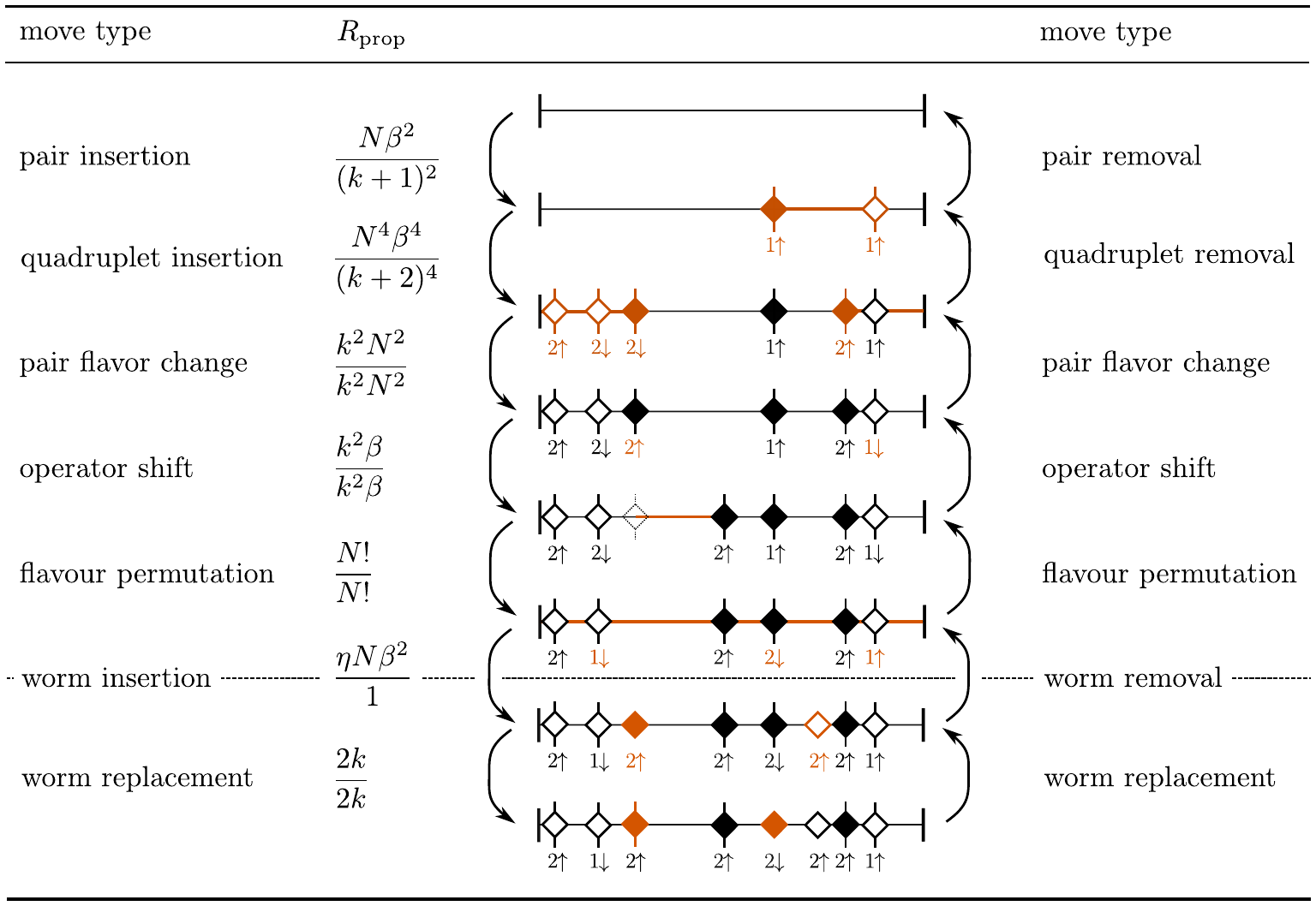}
\par\end{centering}
\caption{Available updates of a configuration in CT-HYB.  The center column
is a diagrammatic representation of the update, where the horizontal line
represents the imaginary time axis $\tau\in[0,\beta]$, filled and empty diamonds
represent local creation and annihilation operators, respectively.  Vertical
lines attached to operators indicate the presence of hybridization lines, i.e.,
operators coming from the hybridization expansion, while the absence of such
lines signifies purely local, i.e. worm, operators. $R_{\rm prop}$ denotes 
a prefactor for the relative probability of the two configurations, which needs to be multiplied to the ratio of the  local and bath  weights discussed in Sections \ref{Sec:LocalTrace} and \ref{Sec:BathTrace}, respectively. Here $N$ is the number of
impurity spin-orbitals, $k$ the expansion order, and $\eta$  the balancing
parameter between worm and partition function space. \label{Tab:CTHYB}
}

\end{table}

The expansion of the thermal expectation value of the partition function
(or equivalently, any Green's function like object) results in an
infinite series of Feynman diagrams which is sampled through the Metropolis Hastings Monte Carlo algorithm.
In the case of CT-HYB this is a series  in terms of hybridization inclusions,
see Table~\ref{Tab:CTHYB}.
 Thus, the Monte Carlo sampling
procedure takes place in the space of all possible Feynman diagrams.
In order to fulfill ergodicity and (detailed) balance, it is necessary
to define Monte Carlo steps, which increase and decrease the expansion
order of the infinite series. We usually relate to these steps as
'add' and 'remove'.

Further Monte Carlo steps allow us to change the observable we sample
(worm sampling) or decrease auto-correlation lengths (global moves
in partition function space and replacement moves in worm space). 

\paragraph{Partition Function Sampling}

When expanding the partition function in the hybridization, the
set of possible diagrams depends on the type of local interaction
considered. The diagrammatic series built from general local interactions
includes Feynman diagrams that are not present in the diagrammatic
series built from  simplified local density-density interactions only.
This is because the imaginary time propagation between creation and annihilation
operators in  Table~\ref{Tab:CTHYB} may then include e.g. a spin flip in two different orbitals. 
Yet, in
both cases the 'add' and 'remove' steps alter the local weight and
the bath weight of the configuration, as discussed  in  Sections \ref{Sec:LocalTrace} and \ref{Sec:BathTrace}, respectively. 

As we employ the sliding-window approach\cite{shinaoka-jsm-2014}, the distance between operators in pair insertions and removals is limited by default. When the maximum $\tau$-difference is not set with the parameter \textsf{TaudiffMax}, an automatic calibration procedure is used to estimate a good value.
The percentage for attempted insertion
and removal steps is implicitly determined by setting the percentages
for attempting all remaining steps.
Fast update
formulas are used in the bath determinant.\cite{gull-rmp-2011,Wallerberger16} 

While 'add' and 'remove' steps alter the diagrammatic series by changing
a small subset of operators only, one can further formulate global
moves, which act on all operators of the current configuration. These are computationally
more expensive, as they require a complete reevaluation of the configuration
weight. In the paramagnetic case, the otherwise explicit symmetrization
over spins may be imposed onto the system by proposing to flip the
spins of all the operators at the same time. This is done for 10 \% of 
the global moves. 
If certain symmetries of the systems
are known beforehand, operator permutations may be specified in the
parameters file with \textsf{SymMove01} to \textsf{SymMove99} and an additional parameter
\textsf{NSymMove} is specifying the number of symmetry moves. 
E.g. for a 5 orbital system \textsf{NSymMove = 1} and \textsf{SymMove01 = 7 2 3 4 5 6 1 8 9 10}
will propose an exchange of orbital 1 spin-up with orbital 2 spin-down (the first 5 numbers indicate
orbitals of spin-up, the last 5 of spin-down).
If \textsf{NSymMove} $\neq 0$, this is done for
fifty percent of the global moves. 
Lastly, random 
permutations of the operator flavours in the trace are performed
for the remaining global moves. 

Global moves are a way to restore (approximate) orbital- or spin-degeneracies in systems where polarized configurations are separated by a large phase space barrier.\cite{PhysRevB.96.155135} They are trivially accepted if they respect an exact symmetry of the system, and become exponentially suppressed as this symmetry is broken.  The total percentage of attempted global moves is specified with the parameter \textsf{PercentageGlobalMoves}, which by default is set to 0.5~\% of  total moves.

\paragraph{Worm Sampling}

Instead of sampling the diagrammatic series of the partition function
and extracting an observable by altering individual configurations,
one may choose to sample an observable directly. Switching from partition
function space (the diagrammatic series given by the denominator of
the thermal expectation value, i.e. importance sampling) to worm space
(the diagrammatic series given by the numerator of the thermal expectation
value) requires a new set of moves. Primarily, one is interested in
extracting Green's function like objects in this fashion, which is
why configuration weights of worm spaces only differ from configuration
weights of the partition function space by a set of additional local
operators (in terms of diagrams these are the external legs). The moves switching
between these two spaces are also of the type 'add' or 'remove', but
here they only alter the local weight, leaving the bath weight unchanged.
The corresponding percentage of attempted inserts and removes is specified
with the parameter \textsf{PercentageWormInsert}. Within worm space, the
sampling with operators connected to the bath is required in order
to fulfill ergodicity. Again, the percentage of attempted inserts
and removes of such operator pairs is implicitly determined by all
remaining steps. While the global moves have not been implemented
in worm space, additional replacement steps exist. These attempt to
change a specific worm configuration by reattaching hybridization
lines from a given operator to the worm operator of the same flavor.
The percentage of attempted replacements is specified by the parameter
\textsf{PercentageWormReplace}.

\subsection{Estimators}
\label{Sec:Estimator}

Once the Monte Carlo sampling of the Feynman diagrammatic series expansion
is established, the remaining task is to extract physical observables.
This procedure differs when sampling the partition function series
and when sampling the observable directly using worm sampling, cf.~Ref.~\onlinecite{gunacker-prb-2015}.

\paragraph{Partition Function Sampling}

The major quantities of interest are the one-particle (imaginary time,
Matsubara and Legendre basis) and the two-particle Green's function
(Matsubara basis). From  the respective Green's
functions, the one-particle and two-particle density matrices
can be obtained as the  all-equal-time components. The histogram 
denotes
the probability distribution of the expansion order $k$, i.e.,  the number of pairs in Figure~\ref{tab:int} (which  in the hybridization expansion depends on the kinetic energy). 
In order to measure an $n$-particle impurity Green's function in {\em Partition Function Sampling} mode, all hybridization lines connected to $n$ operators are removed, in order to create $n$ purely local operators that are disconnected from the bath (also coined ``hybridization sampling''). 
In Table~\ref{Tab:CTHYB} this corresponds to taking away the vertical line of  $n$ filled  creation and $n$ open annihilation operators. 
Bosonic observables like the local DMFT charge susceptibility $\left< n_{i,\sigma}(\tau)n_{j,\sigma'}(0) \right>$ or spin susceptibility $\chi_{\mathrm{loc}}(\tau)$ are measured by insertion of density operators. 
A detailed list of all relevant parameters and the name of the output quantities is given in Table~\ref{tab:est}.

\paragraph{Worm Sampling}

The worm algorithm in continuous-time Quantum Monte Carlo algorithms,
such as CT-HYB, was introduced in order to allow for more flexibility
in terms of the choice of estimators. While partition function sampling
restricts the observables to the diagrams present in the infinite
series expansion of the partition function, this is not true for worm
sampling. Instead, diagrams may be present in the expansion of the observable
itself, which are otherwise absent. This is especially true for the
two-particle Green's function, when considering a diagonal hybridization
function with non-density-density local interactions. Further Green's
functions with a subset of the operators occurring at equal times
can be easily defined for general local interactions. These are important
when extracting the high frequency asymptotics of the fully time-dependent
Green's function. A detailed list of all relevant parameters and the
name of the output quantities is given in Table~\ref{tab:est}.

\begin{table}
\begingroup\sf\small
\begin{tabular}{llr@{\hskip3pt}l}
\toprule 
\addlinespace[0.4em]
output & parameter(s) & \multicolumn{2}{c}{estimator}\tabularnewline
\midrule
\addlinespace[0.4em]
sign & \textendash{} & $s=$ & $\langle\sgn w\rangle_{|w|}$\tabularnewline
\addlinespace[0.4em]
hist & - &  & histogram of expansion order $k$\tabularnewline
\addlinespace[0.4em]
gtau & \textsf{Ntau=$N$} & $G_{i\sigma}(\tau_{k})=$ & $+\frac{1}{\beta}\rect(\frac{\tau_{12}-\tau_{k}}{\beta/N})\langle T_{\tau}c_{i\sigma}(\tau_{1})c_{i\sigma}^{\dagger}(\tau_{2})\rangle$\tabularnewline
\addlinespace[0.4em]
gleg & NLegMax & $G_{i\sigma,l}=$ & $\frac{\sqrt{2l+1}}{\beta}\int_{0}^{\beta}\dd{\tau}P_{l}(\frac{\tau_{12}-\beta}{2\beta})\langle T_{\tau}c_{i\sigma}(\tau_{1})c_{i\sigma}^{\dagger}(\tau_{2})\rangle$\tabularnewline
\addlinespace[0.4em]
giw & \textsf{NGiw,NLegOrder=$L$} & $G_{i\sigma}(\iv)=$ & $\sum_{l=0}^{L}t_{l}(\iv)G_{i\sigma,l}$\tabularnewline
\addlinespace[0.4em]
giw-meas & MeasGiw=1, NGiw & $G_{i\sigma}(\iv)=$ & $-\frac{1}{\beta}\int_{0}^{\beta}\dd{^{2}\tau}\ee^{\iv\tau_{12}}\langle T_{\tau}c_{i\sigma}(\tau_{1})c_{i\sigma}^{\dagger}(\tau_{2})\rangle$\tabularnewline
\addlinespace[0.4em]
rho1 & MeasDensityMatrix=1 & $\rho_{i\sigma,j\sigmap}^{(1)}=$ & $\langle T_{\tau}c_{i\sigma}^{\dagger}c_{j\sigmap}\rangle$\tabularnewline
\addlinespace[0.4em]
rho2 & MeasDensityMatrix=1 & $\rho_{i\sigma,j\sigmap,k\sigmapp,l\sigmappp}^{(2)}=$ & $\langle T_{\tau}c_{i\sigma}^{\dagger}c_{j\sigmap}^{\dagger}c_{k\sigmapp}c_{l\sigmappp}\rangle$\tabularnewline
\addlinespace[0.4em]
sztau-sz0 & MeasSusz=1 & $\chi_{\mathrm{loc}}(\tau)=$ & $g^2 \sum_{ij}  \left< S_z^i(\tau)S_z^j(0) \right>$ \tabularnewline
&  &  & $\hspace{-6em}=g^2  \sum_{ij}  \left< \frac12 (n_{i,\uparrow}(\tau)-n_{i,\downarrow}(\tau))\frac12(n_{j,\uparrow}(0)-n_{j,\downarrow}(0)) \right>$ \tabularnewline
\addlinespace[0.4em]
ntau-n0 & MeasSusz=1 & & $\left< n_{i,\sigma}(\tau)n_{j,\sigma'}(0) \right>$ \tabularnewline
\addlinespace[0.4em]
g4iw & FourPnt=4, & \multicolumn{2}{l}{$G_{i\sigma,j\sigmap,k\sigmapp,l\sigmappp}^{(\mathrm{ph})}(\iv,\ivp,\iw)=\int_{0}^{\beta}\dd{^{4}\tau}\ee^{\iv\tau_{12}+\ivp\tau_{34}+\iw\tau_{14}}$}\tabularnewline
\addlinespace[0.4em]
 & N4iwf,N4iwb &  & $\times\:\langle T_{\tau}c_{i\sigma}(\tau_{1})c_{j\sigmap}^{\dagger}(\tau_{2})c_{k\sigmapp}(\tau_{3})c_{l\sigmappp}^{\dagger}(\tau_{4})\rangle$\tabularnewline\hline
\addlinespace[0.4em]
gtau-worm & WormMeasGtau=1 & $G_{i\sigma,j\sigmap}(\tau_{k})=$ & $+\frac{1}{\beta}\rect(\frac{\tau_{12}-\tau_{k}}{\beta/N})\langle T_{\tau}c_{i\sigma}(\tau_{1})c_{j\sigmap}^{\dagger}(\tau_{2})\rangle_{\!\mathrm{W}}$\tabularnewline
\addlinespace[0.4em]
giw-worm & WormMeasGiw=1 & $G_{i\sigma,j\sigmap}(\iv)=$ & $-\frac{1}{\beta}\int_{0}^{\beta}\dd{^{2}\tau}\ee^{\iv\tau_{12}}\langle T_{\tau}c_{i\sigma}(\tau_{1})c_{j\sigmap}^{\dagger}(\tau_{2})\rangle_{\!\mathrm{W}}$\tabularnewline
\addlinespace[0.4em]
\scalebox{.9}[1.0]{gsigmaiw-worm} & \scalebox{.9}[1.0]{WormMeasGSigmaiw=1} & $[\Sigma G]_{i\sigma,j\sigmap}(\iv)=$ & $-\frac{1}{\beta}\sum_{klm}\sum_{\sigmapp}U_{[ik]lm}\int_{0}^{\beta}\dd{^{2}\tau}\ee^{\iv\tau_{12}}$\tabularnewline
\addlinespace[0.4em]
 &  &  & $\times\:\langle T_{\tau}c_{k\sigmapp}^{\dagger}(\tau_{1})c_{m\sigma}(\tau_{1})c_{l\sigmapp}(\tau_{1})c_{j\sigmap}^{\dagger}(\tau_{2})\rangle$\tabularnewline
\addlinespace[0.4em]
g4iw1-worm & WormMeasG4iw1=1 & \multicolumn{2}{l}{$G_{i\sigma,j\sigmap,k\sigmapp,l\sigmappp}^{(\mathrm{ph})}(\iw)=\int_{0}^{\beta}\dd{^{4}\tau}\ee^{\iw\tau_{14}}$}\tabularnewline
\addlinespace[0.4em]
 & N2iwb &  & $\times\:\langle T_{\tau}c_{i\sigma}(\tau_{1})c_{j\sigmap}^{\dagger}(\tau_{1})c_{k\sigmapp}(\tau_{4})c_{l\sigmappp}^{\dagger}(\tau_{4})\rangle$\tabularnewline
\addlinespace[0.4em]
g4iw1pp-worm & WormMeasG4iw1PP=1 & \multicolumn{2}{l}{$G_{i\sigma,j\sigmap,k\sigmapp,l\sigmappp}^{(\mathrm{pp})}(\iw)=\int_{0}^{\beta}\dd{^{4}\tau}\ee^{\iw\tau_{12}}$}\tabularnewline
\addlinespace[0.4em]
 & N2iwb &  & $\times\:\langle T_{\tau}c_{i\sigma}(\tau_{1})c_{j\sigmap}^{\dagger}(\tau_{2})c_{k\sigmapp}(\tau_{1})c_{l\sigmappp}^{\dagger}(\tau_{2})\rangle$\tabularnewline
\addlinespace[0.4em]
g4iw2-worm & WormMeasG4iw2=1 & \multicolumn{2}{l}{$G_{i\sigma,j\sigmap,k\sigmapp,l\sigmappp}^{(\mathrm{ph})}(\iv,\iw)=\int_{0}^{\beta}\dd{^{4}\tau}\ee^{\iv\tau_{12}+\iw\tau_{23}}$}\tabularnewline
\addlinespace[0.4em]
 & N3iwf,N3iwb &  & $\times\:\langle T_{\tau}c_{i\sigma}(\tau_{1})c_{j\sigmap}^{\dagger}(\tau_{2})c_{k\sigmapp}(\tau_{3})c_{l\sigmappp}^{\dagger}(\tau_{3})\rangle$\tabularnewline
\addlinespace[0.4em]
g4iw2pp-worm & WormMeasG4iw2PP=1 & \multicolumn{2}{l}{$G_{i\sigma,j\sigmap,k\sigmapp,l\sigmappp}^{(\mathrm{pp})}(\iv,\iw)=\int_{0}^{\beta}\dd{^{4}\tau}\ee^{\iv\tau_{12}+\iw\tau_{23}}$}\tabularnewline
\addlinespace[0.4em]
 & N3iwf,N3iwb &  & $\times\:\langle T_{\tau}c_{i\sigma}(\tau_{1})c_{j\sigmap}^{\dagger}(\tau_{3})c_{k\sigmapp}(\tau_{2})c_{l\sigmappp}^{\dagger}(\tau_{3})\rangle$\tabularnewline
\addlinespace[0.4em]
g4iw-worm & WormMeasG4iw=1 & \multicolumn{2}{l}{$G_{i\sigma,j\sigmap,k\sigmapp,l\sigmappp}^{(\mathrm{ph})}(\iv,\ivp,\iw)=\int_{0}^{\beta}\dd{^{4}\tau}\ee^{\iv\tau_{12}+\ivp\tau_{34}+\iw\tau_{14}}$}\tabularnewline
\addlinespace[0.4em]
 & \scalebox{.9}[1.0]{FourPnt=8,N4iwf,N4iwb} &  & $\times\:\langle T_{\tau}c_{i\sigma}(\tau_{1})c_{j\sigmap}^{\dagger}(\tau_{2})c_{k\sigmapp}(\tau_{3})c_{l\sigmappp}^{\dagger}(\tau_{4})\rangle$\tabularnewline
\addlinespace[0.4em]
h4iw-worm & WormMeasH4iw=1 & \multicolumn{2}{l}{$H_{i\sigma,j\sigmap,k\sigmapp,l\sigmappp}^{(\mathrm{ph})}(\iv,\ivp,\iw)=\int_{0}^{\beta}\dd{^{4}\tau}\ee^{\iv\tau_{12}+\ivp\tau_{34}+\iw\tau_{14}}$}\tabularnewline
\addlinespace[0.4em]
 & \scalebox{.9}[1.0]{FourPnt=8,N4iwf,N4iwb} & \multicolumn{2}{r}{$\times\sum_{pqr}\sum_{\sigmapppp}U_{[ip]qr}\langle T_{\tau}\cdag_{p\sigmapppp}\cee_{q\sigma^{\vphantom{\prime}}}\cee_{r\sigmapppp}\cdag_{j\sigmap}\cee_{k\sigmapp}\cdag_{l\sigmappp}\rangle$}\tabularnewline
\bottomrule
\end{tabular}
\endgroup

\caption{Measurable quantities in partition function and worm sampling}
\label{tab:est}
\end{table}

\subsection{Statistical uncertainties}
\label{sec:stat}

Since at its core, CT-HYB is a Monte Carlo procedure, any estimator comes with a statistical uncertainty that drops as $1/\sqrt{N}$ with the number of Monte Carlo measurements $N$ (parameter \textsf{Nmeas}).  The prefactor and thus the choice of $N$ for reaching the desired accuracy depends on configuration space, i.e., the type of system under study, as well as the type of estimator used.  We state a couple of empirical guidelines to aid users:

\begin{enumerate}
\item in partition function space (upper section of Table~\ref{tab:est}), time-independent estimators such as \textsf{rho1} and \textsf{rho2} typically have larger variances than the Green's function \textsf{gtau}, because they are computed from a single point in imaginary time for each diagram.

\item in worm space (lower section of Table~\ref{tab:est}), estimators with more external indices have larger variances, since the components need to be sampled individually.  E.\,g., \textsf{g4iw-worm} is noisier than \textsf{g4iw1-worm}, which in turn is noisier than \textsf{giw-worm}.
\end{enumerate}

Like any Monte Carlo procedure along a Markov chain, CT-HYB is subject to autocorrelation.  This requires each run to be thermalized, i.\,e., a number of initial Monte Carlo steps $N_\mathrm{th}$ must be discarded (parameter \textsf{Nwarmups}). In order to make sure that the Markov chain ``forgets'' its starting point, we need $N_\mathrm{th} > \tau_\mathrm{exp}$, where $\tau_\mathrm{exp}$ is the exponential autocorrelation time.\cite{sokal-lecture-1996}  However, $\tau_\mathrm{exp}$ is usually unknown, which is why one commonly sets $N_\mathrm{th}$ to a fixed fraction, typically 10\% to 50\%,\cite{sokal-lecture-1996,berg-markov-2004} of the total number of Monte Carlo steps.

In order to assess some thermalization effects, the observable \textsf{gtau-mid-step} is provided, which stores the imaginary-time Green's function averaged over the ``center region'' in imaginary time, $\tau \in [0.4 \beta, 0.6 \beta]$, but resolved for each individual Monte Carlo measurement (parameter \textsf{Gtau\_mid\_step=1}).  We empirically found that when this quantity reached a plateau in terms of Monte Carlo steps, the calculation was sufficiently thermalized.

Autocorrelation also means that the number of truly uncorrelated samples for each observable is reduced from $N$ to $N/(1 + 2\tau_\mathrm{int})$, where $\tau_\mathrm{int}$ is the integrated autocorrelation time.\cite{sokal-lecture-1996}  The immediate effect of this is that errorbars, when computed naively over the Markov chain, will be too small.  To avoid this, we compute errorbars from the $N_{\mathrm{cores}}$ different Monte Carlo runs (given by the $N_{\mathrm{cores}}$ CPUs the computation is run on), which are independent provided \textsf{Nwarmups} was chosen large enough. Let $A_i$ be the averaged result on the $i$-th core, the error bars on $A$ are estimated as:
\begin{equation}
   \Delta A^2 = \frac 1 {N_\mathrm{cores}(N_\mathrm{cores} - 1)} \sum_{i=1}^{N_\mathrm{cores}} (A_i - \bar A)^2.
\end{equation}

As many measurements in CT-HYB are expensive, one still wants to measure only every $N_\mathrm{sweep} > \tau_\mathrm{int}$ steps to avoid spending time on measuring cross-correlated results (parameter \textsf{NCorr}).  A reasonable choice for $N_\mathrm{sweep}$ is the average ``renewal time'' of the trace, $\langle k\rangle/R_\mathrm{rem}$, where $k$ is the mean expansion order and $R_\mathrm{rem}$ is the acceptance rate of removing an operator pair.\cite{gull-rmp-2011,Wallerberger16}  Despite these recipes, the actual calculation of the autocorrelation time from the CT-QMC runs would we very useful; doing so for the many different samplings implemented in w2dynamics is cumbersome but an important task for the future.

\subsection{Local trace}
\label{Sec:LocalTrace}

The computation of the local weight (\ref{g_loc_trace}) relates to the solution
and, with each Monte Carlo move, efficient update of an exact diagonalization
problem with a Fock space size of $2^{N_\mathrm{fl}}$.  Apart from possible
conditioning issues of the bath determinant, it is thus easily the computational
bottleneck for a CT-HYB calculation, and several strategies have been developed
to tackle the problem:
matrix-matrix and matrix-vector
codes implement the time-evolution of many-body states in the thermal
expectation value.\cite{gull-rmp-2011} They are defined for any type of local interaction.
The simpler segment representation applicable to density-density interactions 
is thought to be superior due to a favorable scaling in the number of orbitals
and enhanced Monte Carlo acceptance rates.
For an introduction into CT-HYB in its aforementioned variants see Ref.~\onlinecite{gull-rmp-2011}.
The w2dynamics code features the matrix-vector implementation in the eigenbasis of the local Hamiltonian, combined with the newly developed superstate- and state-sampling methods. There the trace (sum) over impurity eigenstates in eq. (\ref{g_loc_trace}) is not performed explicitely, but sampled over in a Monte Carlo procedure.\cite{hausoel2017} Sampling over them in groups defined by the blocks of $H_{\mathrm{loc}}$ (superstate-sampling) is used by default. Sampling each eigenstate individually can be switched on with \textsf{statesampling=1}, which is faster, but may affect the average sign.

Further details on how the local trace is calculated can be found in Ref.~\onlinecite{hausoel2017}.

\subsection{Quantum numbers}\label{sec:qn}

Unless broken by the local non-interacting terms, each type of interaction conserves a set of quantum numbers on the impurity (parameter \textsf{QuantumNumbers}, set by default to the values given in Table~\ref{tab:int}).  In order to be useful in CT-HYB, a quantum number $Q$ not only must commute with $H_{\mathrm{loc}}$, but each annihilation operator $\cee_i$ must map any eigenstate $|q\rangle$ of $Q$ to some other eigenstate $|q^\prime\rangle$.  This corresponds to a block-diagonal form of $H_{\mathrm{loc}}$.  Since the size of the largest block directly enters the exponential scaling, it is important to use as many quantum numbers of the system as possible.

Any spin-independent quartic interaction conserves the number of electrons and the spin in $z$-direction in the system (quantum numbers \textsf{Nt} and \textsf{Szt}).  The Coulomb interaction in a spherical basis, additionally, conserves angular momentum in $z$-direction (quantum number \textsf{Lzt}).  In the case of density-density interaction, the Hamiltonian conserves the number of electon in each spin-orbital (quantum number \textsf{Azt}), i.e. we set for density-density \textsf{QuantumNumbers = Nt Szt Azt}.  When, on the other hand, non-density-density interactions like Slater--Kanamori or the full Hubbard terms are present in $H_{\mathrm{loc}}$, the time-evolution mixes those states. The Kanamori interaction is $\mathrm{SO}(N) \otimes \mathrm{SU}(2)$ symmetric, and has a corresponding conserved quantity (the PS quantum number, in the code called \textsf{Qzt}); thus we have \textsf{QuantumNumbers = Nt Szt Qzt}.

For a general interaction, it is advisable to set \textsf{QuantumNumbers = Nt Szt All}. This choice enables an automatic search for quantum numbers, where the states in the occupation number basis are reordered in such way that $H_{\mathrm{loc}}$ becomes block-diagonal with a minimal block size.

It is worth mentioning that \textsf{QuantumNumbers} in principle can also be used to truncate parts of $H_{\mathrm{loc}}$. This can be used to assess the sensitivity of the result to certain classes of terms in the local Hamiltonian or to pre-converge a DMFT run using a computationally cheaper model.  Its use is intended for expert use only and it is advisable not to use it as an {\em a priori} approximation of the interaction.



Real hopping terms conserving spin in $z$-direction are necessary to treat short range non-local correlations within cluster-DMFT, and real amplitudes breaking $S_z$ for magnetic fields with an $x$-component.
Complex hopping amplitudes are used to describe spin-orbit coupling or magnetic fields with an $y$-component. The real ones are switched on by \textsf{offdiag=1}, the complex ones will soon be available and can be switched on by \textsf{complex=1}.

\subsection{Bath trace}
\label{Sec:BathTrace}

Using an ab-initio Hamiltonian $H(\mathbf{k})$, one usually gets offdiagonal hybridization functions $\Delta(\tau)$ as well as hopping terms $\mu^{\mathrm{imp}}_{\alpha,\beta}\;c^\dagger_\alpha c_\beta$, which can be allowed by setting the parameter \textsf{offdiag=1}.
By rotating the single particle basis these offdiagonal terms can be minimized but never removed for general offdiagonal hybridizations. 

Density-density interactions and offdiagonal hybridization functions can be combined straightforewardly. 
For Kanamori and full-Hubbard interaction combined with offdiagonal hybridization functions complications arise, since interactions such as the spin-flip configuration ($ c_{2\down}^\dagger c_{1\up}^\dagger c_{2\up} c_{1\down} $) or the pair-hopping configuration ($c_{2\down}^\dagger c_{2\up}^\dagger c_{1\up} c_{1\down}$) 
are contained in the configuration space. Their contribution is zero when the hybridization function is diagonal in the flavors. 
They cannot be generated by insertions and removals of pairs of operators, but by insertions or removals of 4 operators \cite{Seth2016}.
Alternatively the spin-flip and pair-hopping configurations could be generated via so called flavorchange-moves, which change the flavour-indices of two operators, e.g. 
$c_{1\down}^\dagger c_{1\up}^\dagger c_{1\up} c_{1\down} \rightarrow c_{2\down}^\dagger c_{1\up}^\dagger c_{2\up} c_{1\down} $. Flavorchange-moves have acceptance rates orders of magnitudes larger than the 4-operator moves, nevertheless we observe more noise with flavorchange-moves compared to 4-operator-moves. 

By default 4-operator-moves are used, and switched on with the parameter \textsf{Percentage4OperatorMove}.

\section{Installation and community site}
\label{Sec:Tutorial}
\subsection{Where to find the project and the code}
\label{Sec:Where}

We have decided to put the source code of this tool developed hitherto in Wien and W\"urzburg (hence the $w2$ in w2dynamics) under the GNU
GPL license Version 3, an established open source license.
This should be the first, but very important step, towards building a larger community\cite{Budd-plos-2015}
around this code. The goal of this community should be to work together
to obtain a DMFT code that is flexible to embed into other
tools via its python interface and flexible with regard to the simulatable
unit cells. 
Following recent trends in the community we have set up 
a github repository at \url{https://github.com/w2dynamics/w2dynamics}
that serves as a central meeting point on where to find the code.
The code is available to the public and people can interact with the developers via the
issue tracking facilities and the wiki pages so that members of our community can collect information
they consider useful for the project.
There is a  public mailing list \url{ w2dynamics-users@list.tuwien.ac.at}
that is also frequented by the developers; and a second developers  mailing
list \url{ w2dynamics-devel@list.tuwien.ac.at}. We
think that it is beneficial to users to become a part of our community
so that they can benefit from the experience of experts in solid state
physics and Monte Carlo techniques. To bring together people from
various backgrounds we plan to organize workshops and community meetings
that everybody is able to attend. Using an open platform like github we have the technical
means to trace the contributions of everybody, to make sure that every
contributor is listed on our project website, or, in the case of greater
contributions, to authorship on a relevant contribution. Organizing
workshops, hosting the servers and investing the time takes some effort
but we are committed to invest that to grow an open community. 

\subsection{Installation}

Since the w2dynamics code base tries to leverage the power of Fortran
and the flexibility of python, the installation is a little bit more
involved. In particular, some dependencies have to be taken care of. We tried
our best to make the installation as simple as possible on as many
platforms as possible. As build system we use cmake and we install
a lot of the required dependencies if they can not be found on the user's system. 
At the time of writing we strictly require
cmake > 2.8.5, 
a FORTRAN90 compatible Fortran compiler,
a C++ compiler that at least exposes the C++11 RNGs since we only provide a trivial fallback,
the BLAS and LAPACK libraries,
a python interpreter version 2.4 or bigger,
the FFTW library,
and an implementation of the MPI API. Without these libraries the installation step will fail.
We will automatically install any of the following packages if it is missing: Those packages are:
the NFFT library,
numpy > 1.2,
libhdf5 > 1.6 with Fortran bindings,
mpi4py,
h5py,
scipy > 0.6 with f2py,
and python-configobj.
 Note that the compilation
of NFFT and libhdf5 necessitates a C compiler and the associated libraries.
Python dependencies that are marked as optional will be downloaded
and locally installed using the pip installer.

Let us assume (i) cmake is installed, (ii) the source files are downloaded, and (iii)
the present working directory is the w2dynamics root folder.
Then the following cmake commands should
give working binaries for the CT-HYB code and the Maxent utility:

\begin{lstlisting}
mkdir build
cd build
cmake ..
make
\end{lstlisting}

We  routinely test the compilation of our code on various linux distributions
such as Ubuntu, Debian, OpenSuse and CentOS. After compilation, we recommend the reader to do the tutorial.
To circumvent the problem of rapidly aging documentation we refer the reader to our wiki page at \url{https://github.com/w2dynamics/w2dynamics/wiki/Tutorials}.
Currently the tutorials show some of the features of w2dynamics.  We show how to calculate the one-particle Green's function, the self-energy and orbital occupations of a one-orbital Bethe lattice Hubbard model infinite dimensions, how to use Wannier Hamiltonians as input, show a benchmark of performing a DMFT run in different basis sets and then comparing the results, and show how to calculate two-particle Greens functions with worm sampling. 

The repository also includes additional test and benchmarks, which serves as a testbed for code development. 
The benchmarks have a structured layout and usually follow the same pattern.
The user has to
\begin{itemize}
\item change to the directory of the benchmark which also contains a short description of the problem ,
\item inspect \textsf{Parameters.in}, which contains the parameters in key/value format (see Table \ref{parameters_table} ),
\item  execute DMFT.py in this working folder.
\item This produces an HDF5 archive with all simulation data (see Table \ref{hdf_table})
\item which can be accessed by the hgrep commands or the ipython notebook.
\end{itemize}

\section{Input/output}\label{sec:io}

\paragraph{Parameters}
The w2dynamics code is mainly controlled by specifying a set of input parameters.
Due to the versatility of the code a large amount of such input parameters exist, while most have reasonable default values pre-defined in 'auxiliaries/confispec'. By supplying a configuration file, these default values can be overwritten with user-defined values. 
A sample config file is distributed together with the code.

The configuration file is structured into the three sections [General], [Atoms] and [QMC], while the [Atoms] section consists of numbered subsections ([[1]],[[2]],...) for each individual atom defined in the unit cell. A detailed list of mandatory user-defined parameters is given in Table~\ref{parameters_table}.

\begin{table}
\begingroup\sf\small
\begin{tabular}{lll}
\toprule
\addlinespace[0.4em]
parameter & option (type) & \multicolumn{1}{l}{description}\tabularnewline
\midrule
\addlinespace[0.4em]
{[General{]}}  &  & \multicolumn{1}{l}{}\tabularnewline
\addlinespace[0.4em]
& ReadIn & wannier90 Hamiltonian (path parameter 'HkFile')\tabularnewline
\addlinespace[0.4em]
& ReadInSO & wannier90 spin-orbit Hamiltonian (path parameter 'HkFile')\tabularnewline
\addlinespace[0.4em]
  DOS & Bethe & Bethe lattice/ semi-circular density of states\tabularnewline
\addlinespace[0.4em]
& EDcheck & discrete bath parameters (input files 'epsk' and 'vk')\tabularnewline
\addlinespace[0.4em]
& nano & finite size (parameter 'readleads' and input file 'leadsfile')\tabularnewline
\addlinespace[0.4em]
& readDelta & hybridization function (input files 'deltatau' and 'deltaiw')\tabularnewline
\addlinespace[0.4em]
 DMFTsteps & (integer) & number of DMFT iterations\tabularnewline
\addlinespace[0.4em]
 beta & (float) & inverse temperature\tabularnewline
\addlinespace[0.4em]
 NAt & (integer) & atoms per unit cell (number of subgroups in {[Atoms{]})}\tabularnewline
\midrule
\addlinespace[0.4em]
{[Atoms{]}}  &  & \multicolumn{1}{l}{}\tabularnewline
\addlinespace[0.4em]
{[{[}1{]}{]} ... {[}{[}NAt{]}{]}} &  & \tabularnewline
\addlinespace[0.4em]
& Density & density-density interaction (defined by Udd, Jdd, Vdd)\tabularnewline
\addlinespace[0.4em]
& Kanamori & Slater-Kanamori interaction (defined by Udd, Jdd, Vdd)\tabularnewline
\addlinespace[0.4em]
 Hamiltonian & Coulomb & Coulomb interaction for the d-shell (defined by F0, F2, F4)\tabularnewline
\addlinespace[0.4em]
& ReadNormalUmatrix & interaction matrix (path specified with 'umatrix')\tabularnewline
\addlinespace[0.4em]
& ReadUmatrix & spin-dependent interaction matrix (path parameter 'umatrix')\tabularnewline
\addlinespace[0.4em]
 Udd & (float) & intra-orbital interaction\tabularnewline
\addlinespace[0.4em]
 Nd & (integer) & number of correlated bands\tabularnewline
\midrule
\addlinespace[0.4em]
{[QMC{]}}  &  & \multicolumn{1}{c}{}\tabularnewline
\addlinespace[0.4em]
 Nmeas & (integer) & Monte Carlo measurement steps\tabularnewline
\addlinespace[0.4em]
 Nwarmups & (integer) & warmup / thermalization steps\tabularnewline
\addlinespace[0.4em]
 NCorr & (integer) & steps between two successive measurement / auto-corrleation \tabularnewline
\bottomrule
\end{tabular}
\endgroup
\caption{ User-defined parameters}
\label{parameters_table}
\end{table}

\paragraph{HDF5 output}
Any quantity measured by the impurity solver or calculated during the self-consistency cycle is stored in hierarchical data format (HDF5).
Nowadays, this format is the de facto standard and is employed by the major DMFT software packages.
File operations for reading and writing data from and to HDF5 are realized through the Python library \textit{h5py}.

The output file consists of several hidden groups ('.config', '.environment', '.axes' and '.quantities'), which store the meta-data of each run.
The '.config' group stores all simulation parameters (user-defined or default values). 
The '.environment' group stores the entire shell-environment and further scheduler-parameters on clusters.
The '.axes' stores Matsubara and imaginary time axes, which are shared between different quantities. 
The link between these quantities and the corresponding axis is defined in '.quantities'.

Apart from the hidden group, the output file consists of several iteration groups ('start', 'dmft-', 'stat-', 'finish').
While 'start' stores information about the non-interacting problem all other iteration groups store quantities further grouped by the inequivalent atoms. A detailed list is given in Table~\ref{hdf_table}.

\begin{table}
\begingroup\sf\small
\begin{tabular}{lllll}
\toprule
\addlinespace[0.4em]
group & parent & child & dataset (attributes)  & \multicolumn{1}{l}{description}\tabularnewline
\midrule
\addlinespace[0.4em]
.config & / & - & (general.dos), ...  & \multicolumn{1}{l}{simulation parameters}\tabularnewline
\addlinespace[0.4em]
.environment & / & - & (SHELL), (PATH), ... & shell environment\tabularnewline
\addlinespace[0.4em]
.axes & / & - & iw, tau, ... & frequency / time axes\tabularnewline
\addlinespace[0.4em]
.quantities & / & giw, gtau, ... & (axes), (desc), ... & metadata for quantities\tabularnewline
\addlinespace[0.4em]
start & / & h-mean, lda-mu, ...  & value & dmft initial data\tabularnewline
\addlinespace[0.4em]
dmft- & / & mu, ineq-, ... & value, error & dmft iteration\tabularnewline
\addlinespace[0.4em]
stat- & / & mu, ineq-, ... & value, error & statistic iteration\tabularnewline
\addlinespace[0.4em]
finish & / & mu, ineq-, ... & value, error & hdf5-link to last iteration\tabularnewline
\addlinespace[0.4em]
ineq- & dmft-, stat- & giw, gtau, ... & value, error & inequivalent atom\tabularnewline
\bottomrule
\end{tabular}
\endgroup
\caption{hdf5-file structure}
\label{hdf_table}
\end{table}

\paragraph{hgrep utility}
For quick access to the data stored in the HDF5 file, the {\tt hgrep} command
line utility is included with the code. hgrep can extract a subset of the data
stored in one or more HDF5 files, and tabulate or plot the resulting data.  As
an example, the following command:
\begin{equation*}
 \texttt{hgrep}
   \underbrace{\texttt{-p}}_\textrm{plot}
   \underbrace{\vphantom{p}\texttt{test.hdf5}}_{\textrm{file}\ }
   \underbrace{\vphantom{p}\texttt{siw}}_{\Sigma(\iv)\ }
   \underbrace{\vphantom{p}\texttt{1:3}}_\textrm{iterations\ }
   \underbrace{\vphantom{p}\texttt{1}}_{\textrm{atom }(I)\ }
   \underbrace{\vphantom{p}\texttt{1,4}}_{\textrm{orbital }j\ }
   \underbrace{\vphantom{p}\texttt{1}}_{\textrm{spin }\sigma\ }
   \underbrace{\vphantom{p}\texttt{0:20}}_{\iv}\
   \underbrace{\vphantom{p}\texttt{field=value-im}}_{\textrm{only imaginary part}} 
\end{equation*}
plots the imaginary part of the impurity self-energy stored in the first {\tt test.hdf5} for the first three iterations (FORTRAN-style
ranges are supported), impurity problem $I=1$, orbitals 1 and 4, spin up, in
the frequency range $0\le\iv\le20$.  A man page is available, which lists
more detailled syntax and provides further examples.

\section{Conclusions}
\label{Sec:conclusion}

We have described here the main features of the hybridization-expansion continuous-time quantum Monte Carlo code package ``w2dynamics''.
Users can calculate local two- and four-point fermionic Green's functions of multi-orbital impurity models.
These can be either isolated impurities attached to leads or auxiliary impurities in  DMFT or DFT+DMFT calculations of lattice Hamiltonians. The level of approximation is the DMFT; w2dynamics does not introduce further
approximations to the numerically exact CT-QMC algorithm (as e.g.\  inner or outer truncations of the local trace).

The w2dynamics package is similar in spirit to TRIQS \cite{TRIQS,Seth2016,TRIQS/DFTTools} and ALPS \cite{ALPS2}, all of which have their particular strengths.
In the case of w2dynamics these are:
\begin{enumerate}
 \item The calculation of arbitrary one- and two-particle Green's functions and local physical susceptibilities. This serves as an input for subsequent calculations of  physical susceptibility and for diagrammatic extensions of DMFT\cite{RMPVertex} that use  the local vertex as a building block.
The package directly interfaces with  post-DMFT packages such as AbinitioDGA \cite{Galler2017b} and LadderDGA.\cite{Rohringer2018} 

\item Improved sampling techniques such as worm and superstate sampling.   Worm sampling\cite{gunacker-prb-2015} overcomes  the
limitations of the standard hybridization measurement in partition function sampling,
and allows to directly estimate {\em all components} of the $n$-particle Green's function  including equal-time
dynamic susceptibilities, improved estimators for the self-energy,\cite{gunacker-prb-2015} and asymptotics of the four-point vertex.\cite{Kaufmann2017} Moreover, it extends the applicability of the CT-HYB algorithm
 to strongly insulating multi-orbital systems, where partition function sampling
 suffers from severe
ergodicity problems.\cite{gunacker-prb-2015}  Superstate sampling in turn allows to speed up the computation of the local weight by sampling over the block structure of the local Hamiltonian.\cite{hausoel2017}

\item A large versatility of  four-fermion interactions, ranging from the simplest density-density, to the rotationally-invariant Kanamori  to the most general full Coulomb four-index tensor. 
Retardation effects (such as coming from electron-phonon interaction or cRPA screening) can be taken into account.\cite{DanielAndiUw}

\item  A flexible way of interfacing to DFT calculations also for materials with large unit cells made of atoms that can be inequivalent regarding symmetry, number of correlated orbitals, number of ligands, values of the Coulomb interaction, spin orientation, and so on
The input of wannier90\cite{Mostofi2008} generated Hamiltonians, after a transformation to $k$-space, is possible.

\item Treating magnetic phases and off-diagonal  hybridization functions which may stem e.g.\  from (real-valued) inter-orbital hoppings. Future releases will offer full support for complex inter-orbital and inter-spin elements in order to run calculations with spin-orbit coupling.
\end{enumerate}

\subsection*{Acknowledgments}
We are indebted to M.~Ferrero for his help in the early stages of this project
and to  B.~Hartl, O.~Janson, J.~Kaufmann, P.~Chalupa, M.~Sch\"uler, D.~Springer and A.~Valli 
 for contributions to the development of the w2dynamics package. 
We thank these as well as E.~Gull, P.~Werner, A.~Toschi, J.~Tomczak, J.~Kune\v{s}, D.~Di Sante, M.~Aichhorn, A.~Hariki for helpful discussions.  
Financial support is acknowledged from  the Austrian Science Fund (FWF) 
SFB ViCoM F41 (MW), the Deutsche Forschungsgemeinschaft (DFG) research unit
through  SFB1170 (AH,AK,FG,GS) and  research unit FOR 1346 (AH,KH,GS),  the Vienna Scientific  Cluster  (VSC)  Research  
Center  funded  by  the Austrian Federal Ministry of Science (PG),
and the  European Research Council under the European Union's Seventh Framework Program (FP/2007-2013) through ERC grant n.\ 306447 (KH).
AH,AK,FG and GS gratefully acknowledge the Gauss Centre for Supercomputing e.V. for funding this project by providing computing time on the GCS Supercomputer SuperMUC at Leibniz Supercomputing Centre (LRZ).

\end{document}